\documentclass{article}

\usepackage{arxiv}

\usepackage[utf8]{inputenc} % allow utf-8 input
\usepackage[T1]{fontenc}    % use 8-bit T1 fonts
\usepackage{hyperref}       % hyperlinks
\usepackage{url}            % simple URL typesetting
\usepackage{booktabs}       % professional-quality tables
\usepackage{amsfonts}       % blackboard math symbols
\usepackage{nicefrac}       % compact symbols for 1/2, etc.
\usepackage{microtype}      % microtypography
\usepackage{lipsum}		% Can be removed after putting your text content
\usepackage{graphicx}
\usepackage{natbib}
\usepackage{doi}

\usepackage[dvipsnames]{xcolor}

\title{Could mass eccentricity explain the formation of orbits in wind turbines?}

\author{ \href{https://orcid.org/0000-0001-8717-9688}{\includegraphics[scale=0.06]{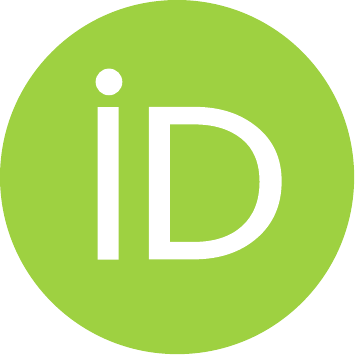}\hspace{1mm} Aljoscha Sander} \\
	Energy and Sustainability Research Institute Groningen\\
	University of Groningen\\
	Groningen, the Netherlands \\
	Institute for Integrated Product Development \\
	University of Bremen\\
	Bremen, Germany\\
	\texttt{aljoscha.sander@rug.nl} \\
	
	\And
	Andreas Haselsteiner \\
	University of Bremen\\
	Institute for Integrated Product Development\\
	ForWind - Center for Wind Energy Research\\
	Bremen, Germany\\
	\texttt{a.haselsteiner@uni-bremen.de} \\
	\And
	Bas Holman \\
	\texttt{b.holman@outlook.com}\\
}

\renewcommand{\shorttitle}{Mass eccentricity in offshore wind turbines}

\hypersetup{
pdftitle={Could mass eccentricity explain the formation of orbits in wind turbines?},
pdfsubject={},
pdfauthor={Aljoscha Sander, Andreas Haselsteiner, Bas Holman},
pdfkeywords={Offshore Wind, Offshore Wind Turbine Installation, Structural Dynamics},
}

\begin{document}
\maketitle

\begin{abstract}

The kinematics of offshore wind turbines are of great importance when installing the turbines, as the motions of the components during craning operations are a limiting factor. Most critical is the installation of the blades: the blade's bolts need to be inserted into the rotor flange, an operation that requires great precision. Both the blade and the turbine undergo environmental loading, leading to relative motions between the blade root and the hub during installation.
Results from an offshore wind farm installation measurement campaign showed, that the partially installed turbines show intricate patterns of motion (orbits) in the horizontal plane. The mechanism behind the formation of these orbits remains elusive so far.

In this paper, we present a novel torsional coupling mechanism linking motions in the fore-aft and side-side direction. It can explain the formation of orbits that change direction. 

\end{abstract}

% keywords can be removed

\section{Introduction}
\label{sec:introduction}

Offshore wind is growing continuously and is on its way to become one of the pillars of green energy. Recent cost reductions in offshore wind (29\% between 2010 and 2019) have mostly been due to increasing turbine size \citep{irenaRenewablePowerGeneration2020}. Increasing turbine size, however, leads to increasing difficulties during installation: larger components are more challenging to handle and require more space on an installation vessel. Today, most turbines are installed component by component, including the blades, a process generally referred to as single blade installation \citep{jiangInstallationOffshoreWind2021}. Blades are installed utilizing a specialized tool, referred to as an installation yoke, that securely fastens the blade during craning operations. Once the blade reaches hub height, a guiding pin is used to land the blade's bolts in the corresponding holes in the rotor flange. During this operation, relative motions between the blade root and the rotor hub greatly hamper the secure landing of the blade in the rotor's flange. As these relative motions are driven by environmental loading on both the turbine and the blade, they are difficult to understand and even more so to predict. 

\clearpage

\begin{figure}
    \centering
    \includegraphics[width=0.7\linewidth]{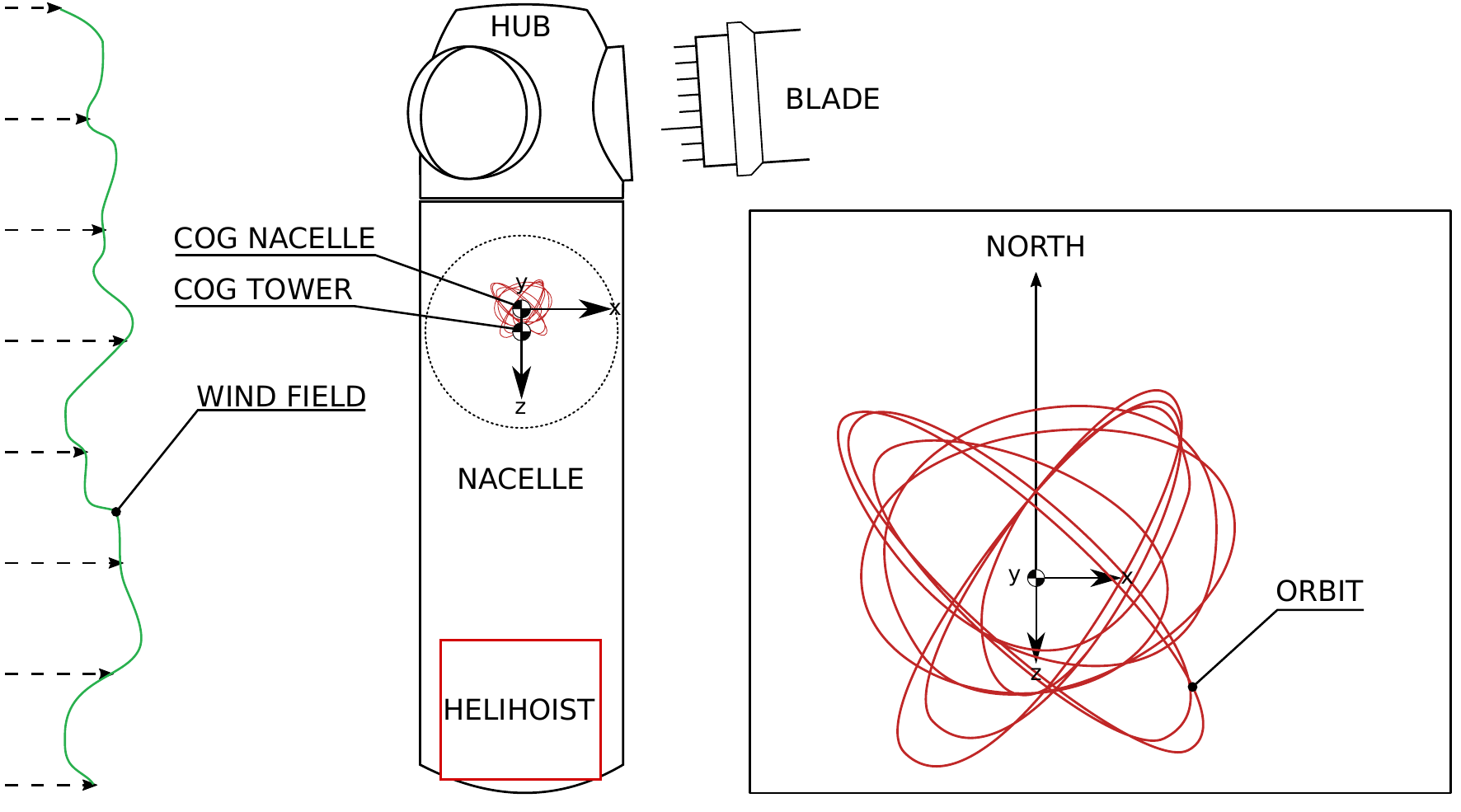}
    \caption{Formation of orbits as observed during the installation of the offshore wind park ``Trianel Windpark Borkum II.'' Note: to enhance visibility, the orbits are not to scale.}
    \label{fig:orbit}
\end{figure}

In a measurement campaign conducted during the installation of the offshore wind farm ``Trianel Windpark Borkum II'' off the coast of Germany in the German Bight, the kinematics of blades and partially installed turbines were investigated \citep{sanderRelativeMotionSingle2020,  sanderMONITORINGOFFSHOREWIND2020, sanderOscillationsOffshoreWind2020}. These measurements revealed, that the partially installed turbines, consisting of a monopile foundation, transition piece, the tower and the nacelle, were displaying intricate patterns of motions — orbits — during single blade installation. A two-minute orbit captured during the installation of the wind park Trianel Windpark Borkum II is shown in \autoref{fig:orbit}. In these orbits, amplitude and direction of motion change swiftly. It stands to reason that these orbits influence the installation of blades and may even be responsible for an increase in installation difficulty.

Several mechanisms may in theory explain the formation of orbits. In this paper, we present how mechanical coupling between the fore-aft and side-side first motion leads to the formation of orbits in a cantilevered beam with an eccentric mass present at the free end. 

\begin{figure}
    \centering
    \includegraphics[width=0.7\linewidth]{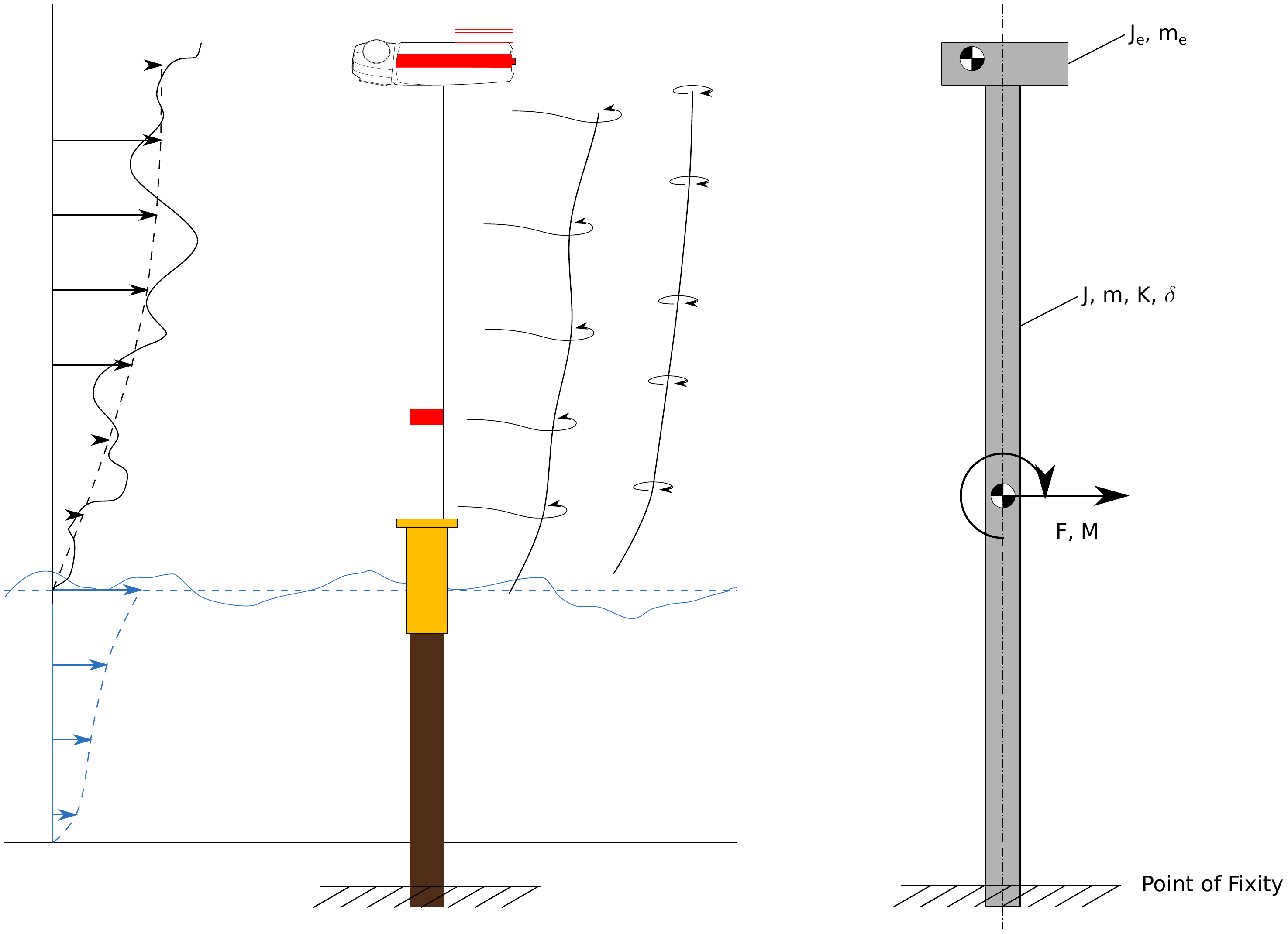}
    \caption{Hammerhead configuration (left) and simplified mechanical model of the hammerhead configuration (right) as a cantilevered beam with an eccentric top mass $m_e$. Wind and wave forces induce a loading force and moment $F, M$. The cantilevered beam has a mass $m$, a moment of inertia $J$, an elasticity $K$, and a structural damping $\delta$. The eccentric mass at the beam's free end also has a moment of inertia $J_e$ }
    \label{fig:loading}
\end{figure}

We focus on one specific state of an offshore wind turbine during installation: the hammerhead configuration. In this configuration, the foundation, transition piece, tower, and nacelle have already been installed, and the next installation step would be the installation of the first blade. In this configuration, the turbine is top-heavy and, due to the nacelle's and rotor's weight, also nose-heavy. As for the turbines installed at Trianel Windpark Borkum II, the centre of gravity of the nacelle is approximately 0.3\,m shifted towards the rotor with respect to the tower axis. \autoref{fig:loading} depicts the hammerhead configuration as well as the simplified mechanical system we derive to investigate the behavior of partially installed turbines during installation. 

A partially installed offshore wind turbine in the hammerhead configuration resembles a cantilevered beam with an eccentric mass at the free end. There are two main directions when viewing the cantilevered beam from above: Fore-aft (North-South) and side-side (East-West). Consider that the eccentric mass is located north of the tower center on the fore-aft axis. If the beam undergoes lateral vibrations in the side-side direction the inertia of the eccentric mass leads to torsion of the cantilevered beam.

This twisting of the beam along its main axis has two effects: 

\begin{itemize}
    \item[a)] The eccentric mass starts to oscillate around the tower center, effectively turning into a torsional oscillator
    
    \item[b)] Due to the circular motion of the eccentric mass around the tower center, transfer of momentum from the initial direction of motion — side-side — to the second direction — fore-aft — takes place.
\end{itemize}

\begin{figure}
    \centering
    \includegraphics[width=0.7\linewidth]{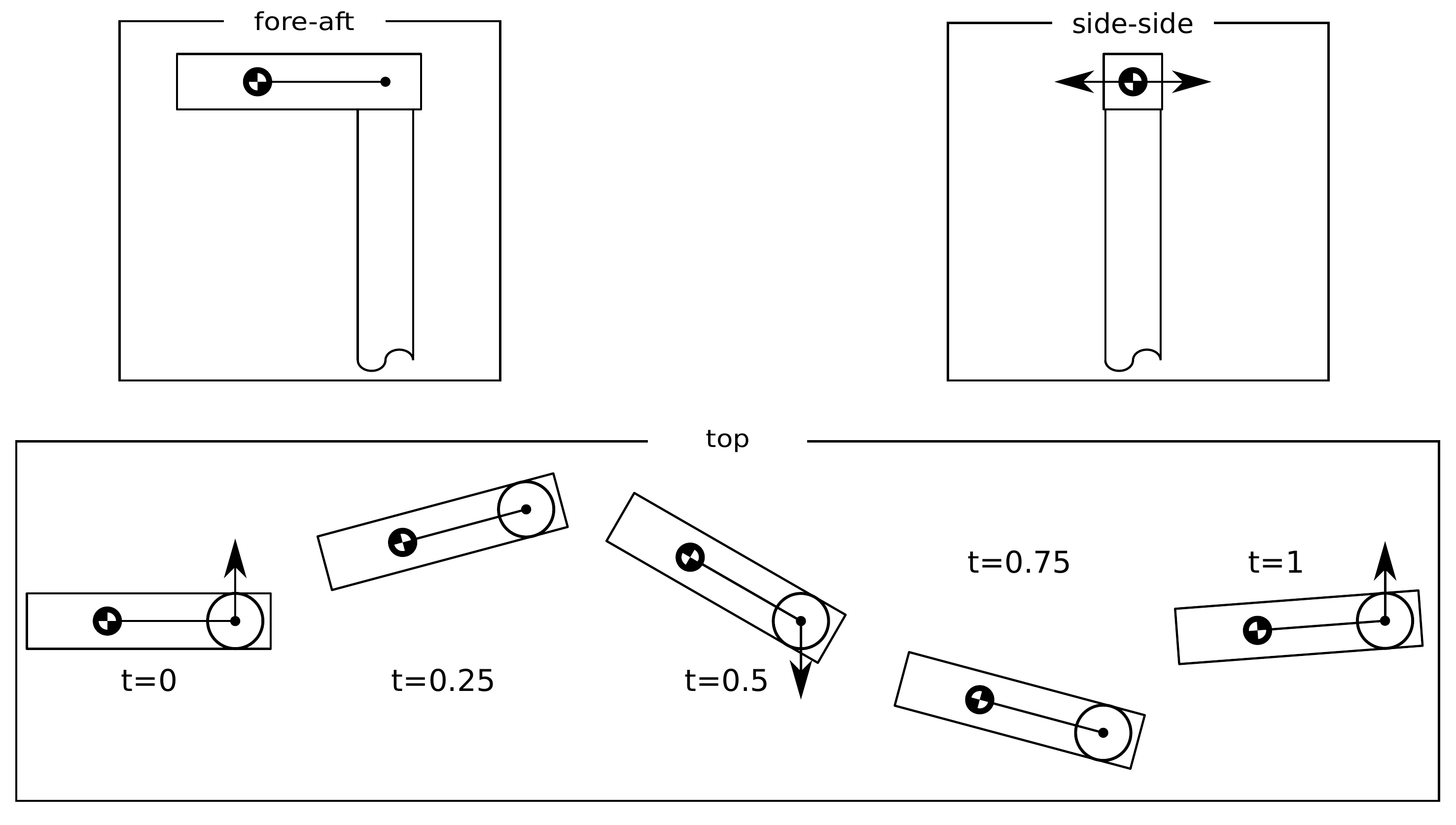}
    \caption{Simplified kinematics of a cantilevered beam with an eccentric mass at the free end undergoing harmonic oscillations in the side-side direction. When viewed from above, the inertia of the eccentric mass induced rotation around the axis of the beam, leading to transfer of momentum.}
    \label{fig:kinematics}
\end{figure}

The torsional oscillation of the eccentric mass in illustrated in \autoref{fig:kinematics}.

In this paper, we set out to investigate this effect of torsional coupling with a table-top experiment, finite element simulations, and a reduced order model approach. 

The paper is subdivided into four sections: First, results from a table-top experiment (\autoref{sec:experiment}), capable of producing orbits that change direction and shape over time, are shown. A section on finite element simulations (\autoref{sec:simulations}) of a comparable system follows. Finally, a discrete system of two bodies, described with differential equations (\autoref{sec:3dof}) is presented. The final section (\autoref{sec:discussion}) provides a discussion and suggests future research into the topic. 

\clearpage

\section{Table-Top Experiment: Cantilevered Beam with an Eccentric Mass}
\label{sec:experiment}

\subsection{Experimental Setup}

As a first proof of concept, we built a table-top experiment that allows us to demonstrate the effect of an eccentric mass on the vibration of a cantilevered beam. As a beam, we chose a wooden stick (massive beech, 0.006\,m diameter) with a free length of 0.7\,m. At the free end of the beam, we placed a 3D-printed lever that allowed us to add M8 bolts (0.028\,m length, 0.0129\,kg each) to control the amount of eccentric mass. Lever length was 0.082\,m weighing a total of 0.026\,kg. The lever was oriented, such that the eccentric mass was positioned north of the beam's axis in the fore-aft plane. The side-side direction is thus defined as perpendicular to the fore-aft direction. 

\begin{figure}[ht!]
    \centering
    \includegraphics[width=0.5\linewidth]{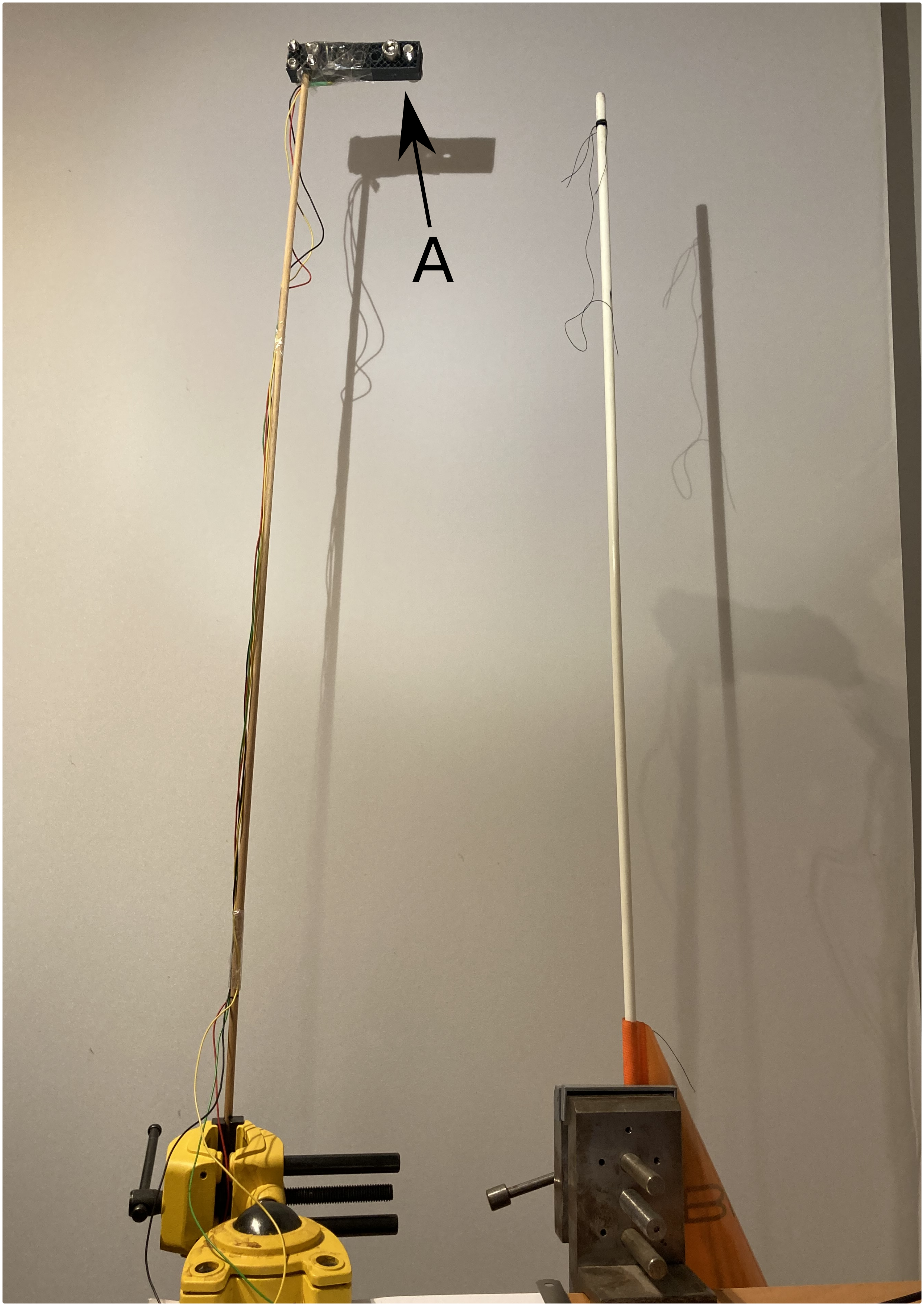}
    \caption{Experimental setup of the table-top experiment. (A) indicates the eccentric mass, added to the cantilevered beam via a 3D-printed lever.}
    \label{fig:setup}
\end{figure}

A second beam made of fibre glass (diameter 0.008 m) was placed next to the cantilevered beam. \autoref{fig:setup} shows the experimental setup. The second beam was used to displace the cantilevered beam from its resting position by pulling the cantilevered beam towards the second beam using a thin cotton thread. The initial displacement of the top of the beam was 0.135 m and three different masses (no bolts, two bolts, four bolts) were tested in the experiment. To release the cantilevered beam from its initial, displaced position, the thread was set on fire using a lighter. For each mass, the experiment was repeated three times. A MEMS-based inertial measurement unit (MPU 9255, InvenSense TDK, Tokyo, Japan) placed on top of the cantilevered beam measured linear acceleration and angular velocity with a sampling rate of 100\,Hz. Measurements were recorded using a Raspberry Pi 400 and analyzed using a jupyter notebook with python3.

We applied the following post-processing steps to the recorded data:

First, obtained accelerations in fore-aft and side-side direction were resampled to 50 Hz using linear interpolation. As the data is indexed by the time of the recording, the time series of each experiment had to be normalized to enable comparison of the different time series. The first peak in the fore-aft acceleration signal in each recorded experiment served as a normalization time stamp; each signal was cropped, such that it starts 0.5\,s prior to the first peak and lasts for 120\,s. 

Following time normalization, the data was filtered by applying a Butterworth high-pass filter with a cut-off frequency of 0.5\,Hz, order 11 and padding of 25\,s to remove any transient effects. The filter was applied forwards and backwards in time to avoid introducing an artificial phase delay. Afterwards, the accelerations and the angular velocities were integrated using a second order trapeze type integration, yielding the velocity vector of the beam's tip and the torsion angle of the beam. To obtain the displacement vector, the velocity vector was filtered again with the same Butterworth filter, integrated again and finally both torsion angle and displacement vector were filtered one last time with the same Butterworth filter.

To compare experimental measurements with finite element simulations and the 3DOF model (both without damping), the observed displacement from the experimental results was normalized to remove the influence of damping from the observed orbits. To this end, the observed decay in displacement amplitude over time is estimated using an exponential function of the following form:

\begin{equation}
    A = A_0 e ^ {\gamma t} + a
\end{equation}

where $A_0$ is the initial amplitude, $\gamma$ is the damping coefficient, and $a$ is a constant offset to account for the finite time series. To obtain orbits free of damping, the displacement vector is scaled by the exponential decay function. 

\subsection{Experimental Results}

\begin{figure*}[ht!]

    \centering
    \begin{subfigure}[b]{0.45\textwidth}
        \centering
        \includegraphics[width=\textwidth]{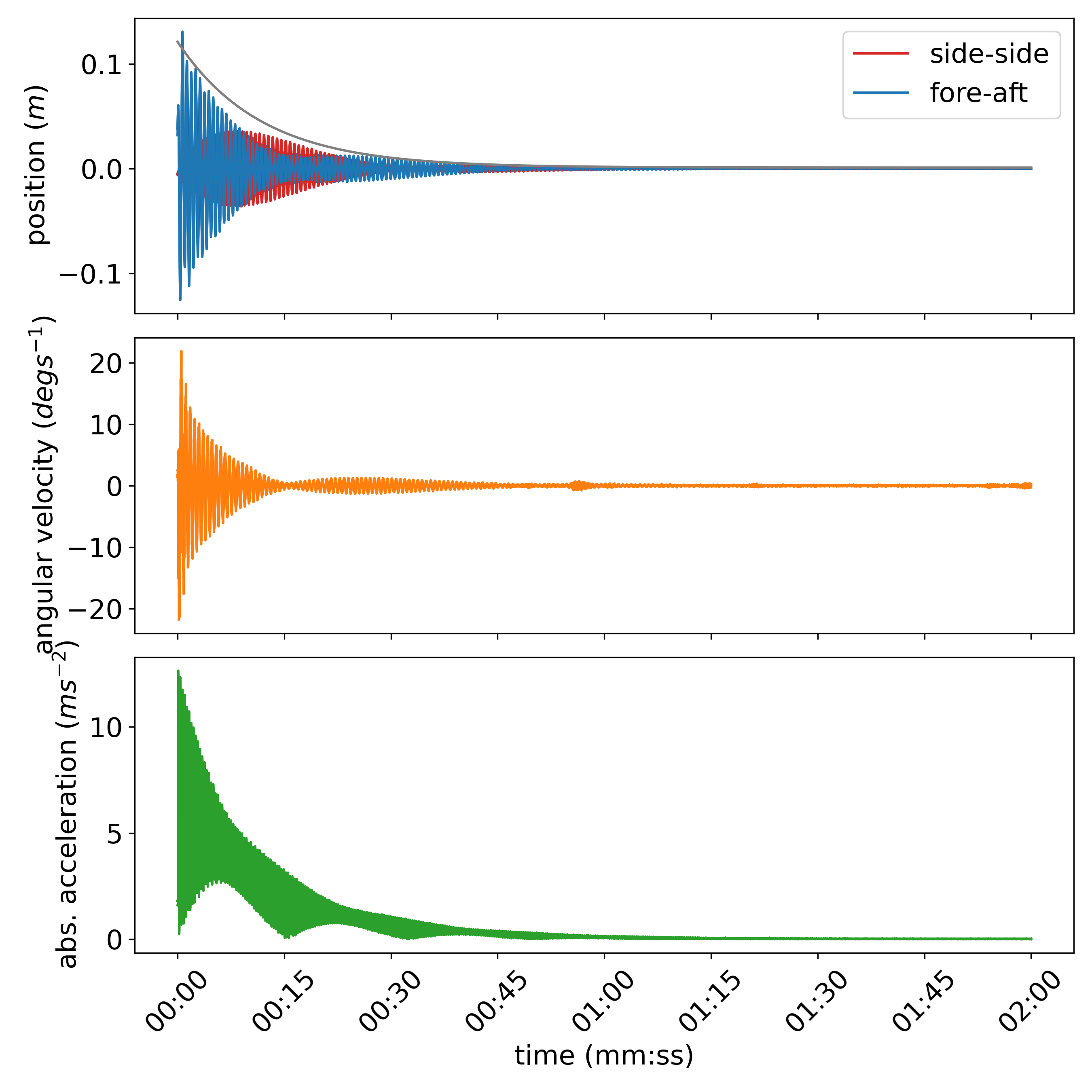}
        %\caption{}
        %\label{fig:low-mass:acc}
    \end{subfigure}
    \begin{subfigure}[b]{0.45\textwidth}
        \centering
        \includegraphics[width=\textwidth]{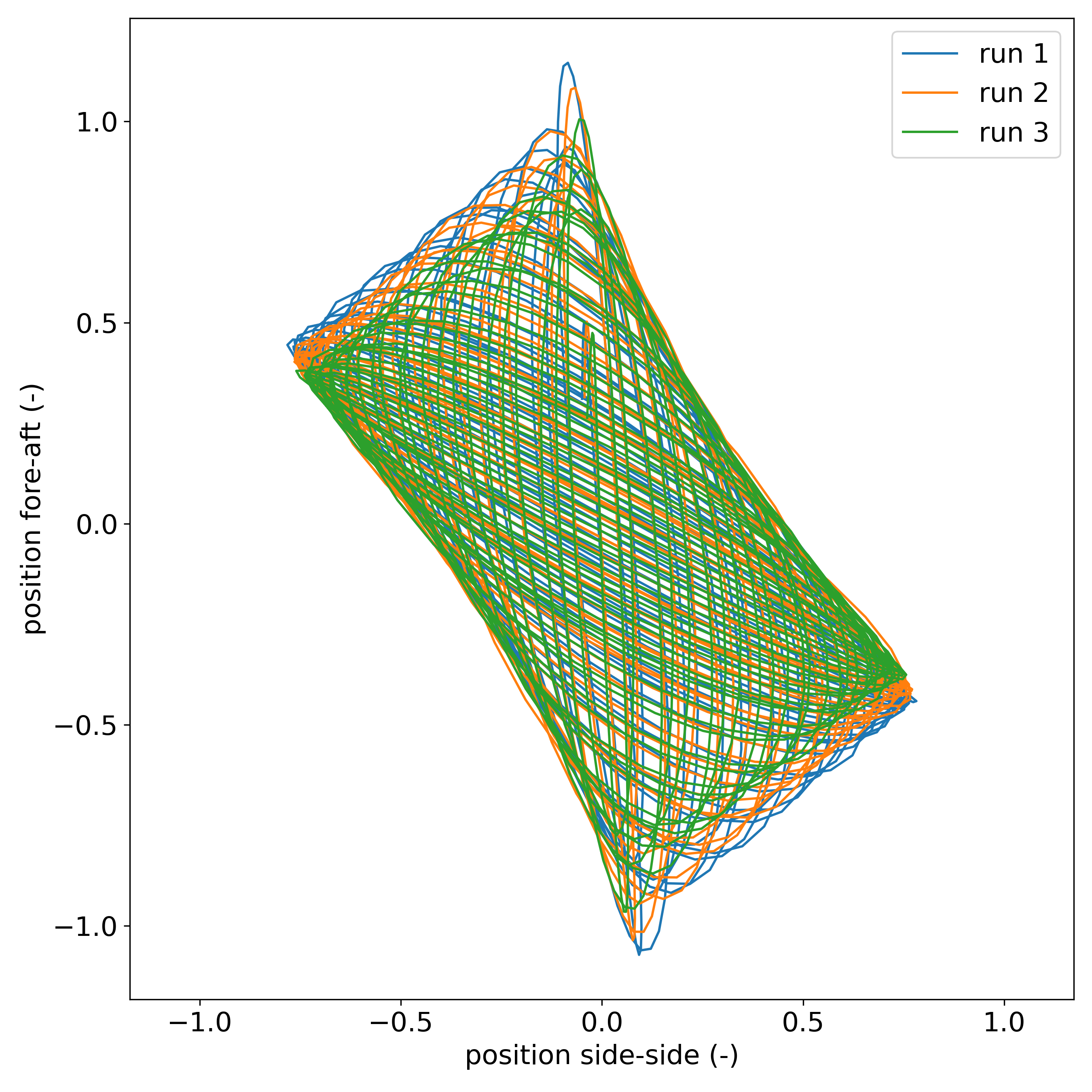}
        %\caption{}
        %\label{fig:low-mass:orbit}
    \end{subfigure}
    
    \caption{Position, angular velocity, and absolute acceleration (left) and Lissajous figures (orbits; right) for the low mass scenario (right).}
    \label{fig:low-mass}
\end{figure*}

\begin{figure*}

    \centering
    \begin{subfigure}[b]{0.45\textwidth}
        \centering
        \includegraphics[width=\textwidth]{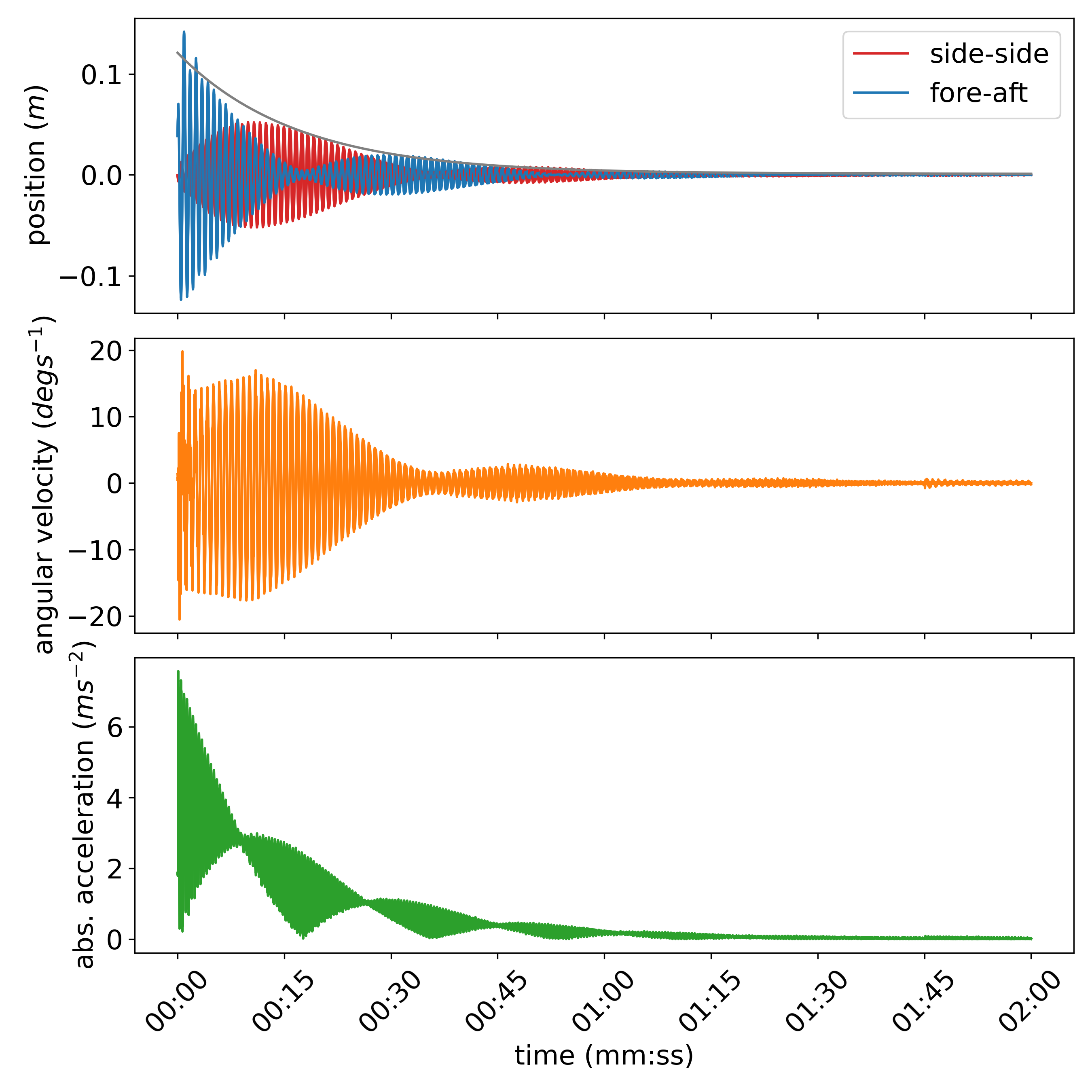}
        %\caption{}
        %\label{fig:medium-mass:acc}
    \end{subfigure}
    \begin{subfigure}[b]{0.45\textwidth}
        \centering
        \includegraphics[width=\textwidth]{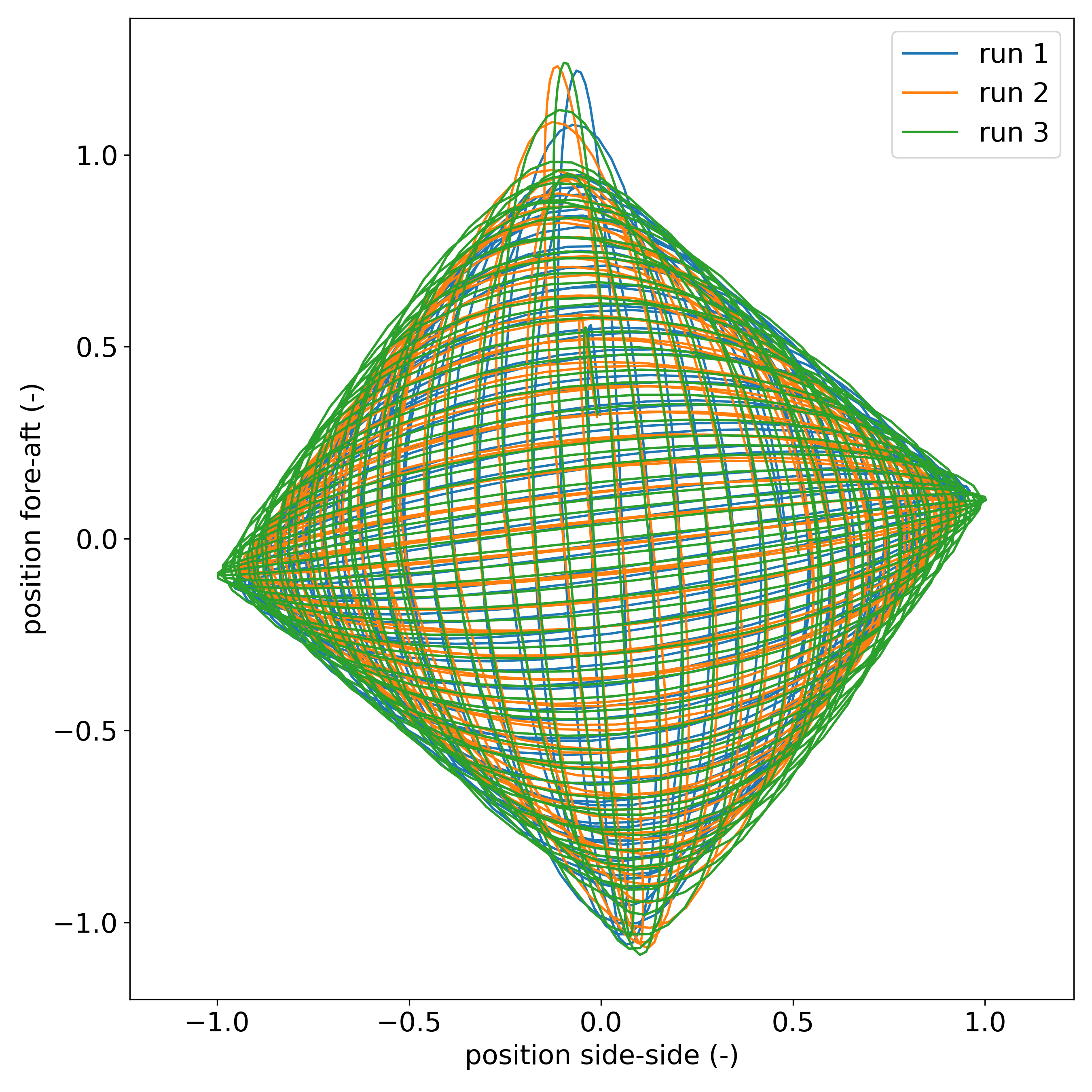}
        %\caption{}
        %\label{fig:medium-mass:orbit}
    \end{subfigure}
    
    \caption{Accelerations, angular velocity and absolute acceleration (left) and Lissajous figures (orbits; right) for the medium mass scenario.}
    \label{fig:medium-mass}
\end{figure*}

\begin{figure*}

    \centering
    \begin{subfigure}[b]{0.45\textwidth}
        \centering
        \includegraphics[width=\textwidth]{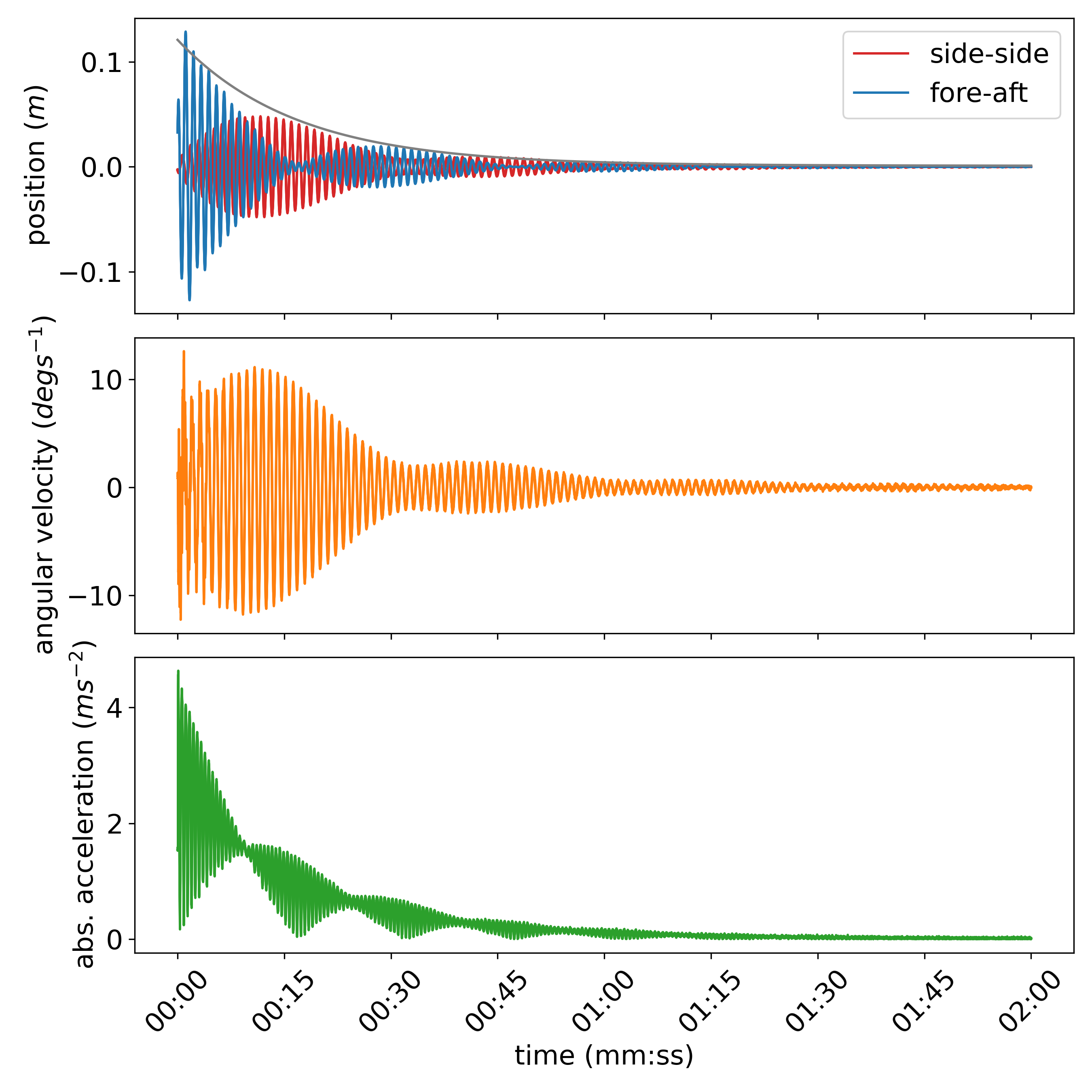}
        %\caption{}
        %\label{fig:high-mass:acc}
    \end{subfigure}
    \begin{subfigure}[b]{0.45\textwidth}
        \centering
         \includegraphics[width=\textwidth]{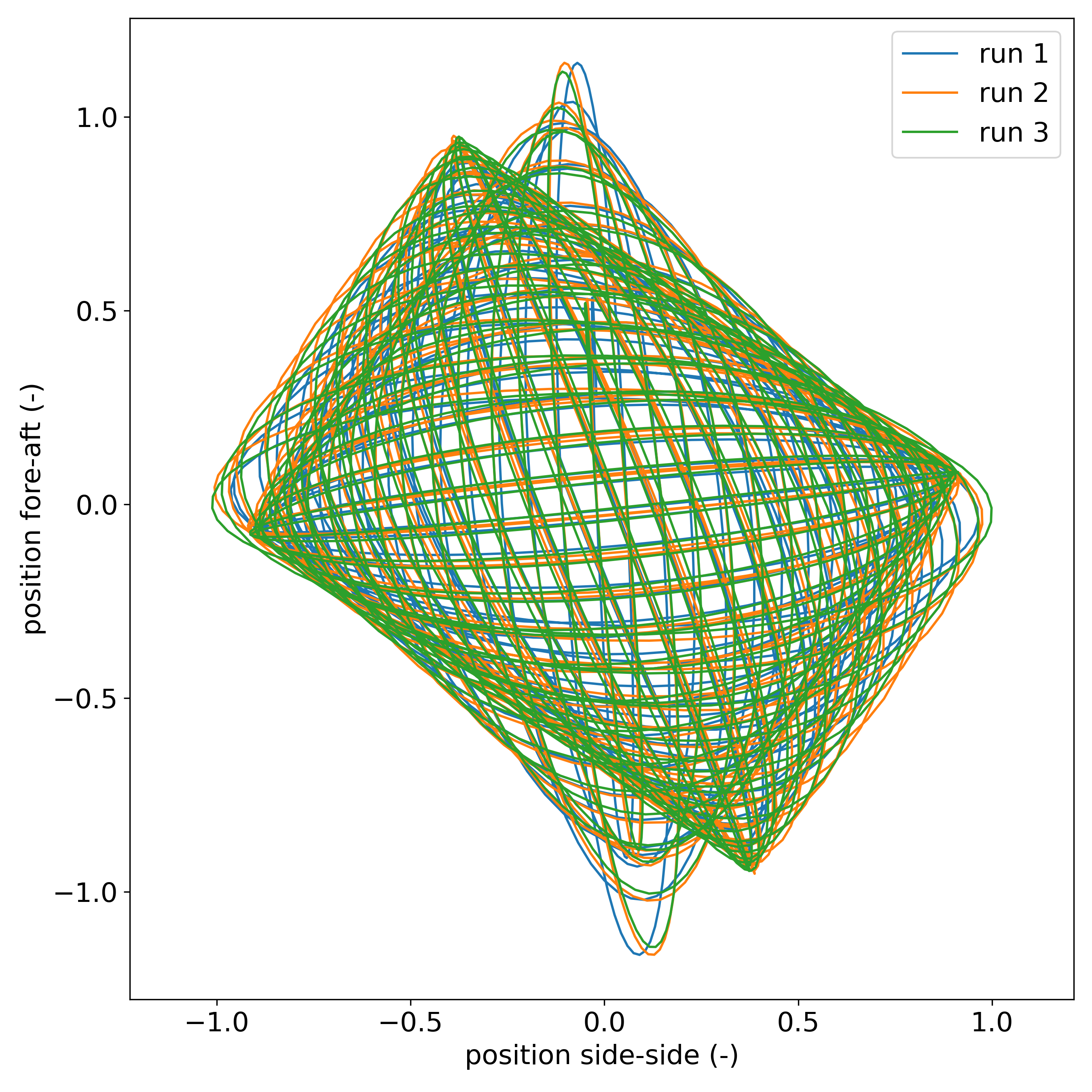}
        %\caption{}
        %\label{fig:high-mass:orbit}
    \end{subfigure}

    \caption{Accelerations, angular velocity and absolute acceleration (left) and Lissajous figures (orbits; right) for the high mass scenario.}
    \label{fig:high-mass}
\end{figure*}

\autoref{fig:low-mass}, \autoref{fig:medium-mass}, and \autoref{fig:high-mass} each show the calculated position as a function of time, the angular velocity along the beam main axis and the absolute acceleration in the fore-aft-side-side plane as well as the corresponding orbits. 

%%%
% Low mass
%%%

%%%
% medium mass
%%%

%%%
% High mass
%%%

The area enclosing all observed orbits seems to be a function of the eccentric mass. In the case of the medium mass scenario, the area is almost quadratic, whereas for the low and high mass scenarios the area is more distorted. For each mass scenario, three runs of the experiment are shown, and the observed orbits are in close resemblance between consecutive runs. With increasing mass, decay is slower. Furthermore, for the medium and high mass scenarios, the angular velocity around the beam axis increases after the initial release and torsion seems to last longer than in the low mass scenario. In all three mass scenarios, torsion of the eccentric mass around the beam's main axis can be observed. It seems that, as long as torsion takes place, changes in direction - and thus orbits - are present.

\section{Finite Element Simulations of a Cantilevered Beam with an Eccentric Mass}

\label{sec:simulations}

\subsection{Simulation Setup}

\begin{figure}[ht!]
    \centering
    \includegraphics[width=0.15\linewidth]{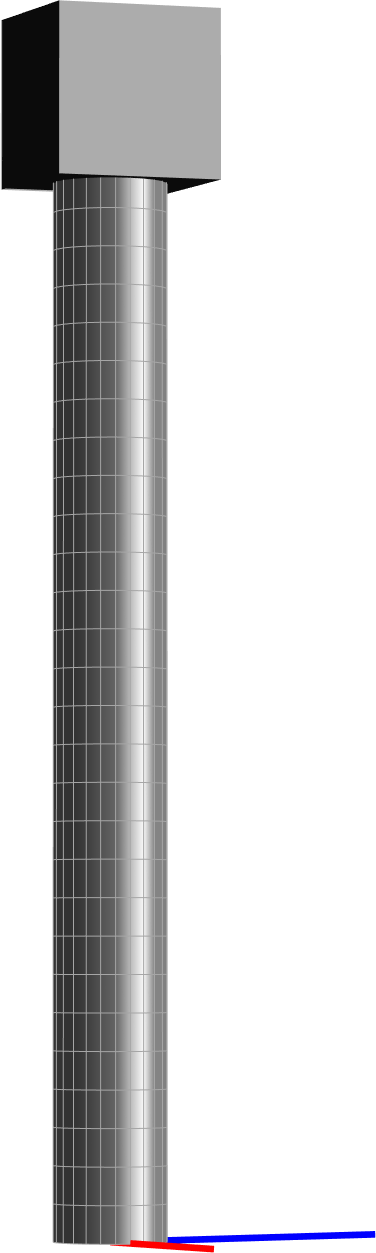}
    \caption{FE model.}
    \label{fig:fea:model}
\end{figure}

A numerical model based on the simplified system shown in \autoref{fig:loading} was created using MATLAB, presented in \autoref{fig:fea:model}. The cantilevered beam was modeled as a hollow cylinder made of  homogeneous linear elastic material (structural steel s235) and with a constant cross-section (diameter: 1.5\,m, wall thickness: 15 mm) along its height (30 m), supported by linear uncoupled springs (\autoref{eq:fea:K_fnd}). The nacelle was modeled as a concentrated mass ($\mathbf{m_{t}}$) with inertia and an additional eccentric mass in both the horizontal and the vertical plane, closely resembling the table-top experiment.

The equations of motion of the undamped system are

\begin{equation}
    \mathbf{M\ddot{\mathbf{u}}}+\mathbf{K}\mathbf{u} = \mathbf{f},
    \label{eq:fea:system}
\end{equation}

where $\mathbf{u} \in \mathbb{R}^6$ describes the concentrated mass' translational ($x,y,z$) and rotational ($\theta_1, \theta_2, \theta_3$) position as a column vector.

The equations were  obtained by superposition of the equations of motions of the concentrated mass ($\mathbf{M}_{top}$) at the free end of the cantilevered beam, the equations for the support springs and the equations for the elasticity of the cantilevered beam itself ($\mathbf{K} = \mathbf{K_{mt}} + \mathbf{K_{found}}+\mathbf{K}_{beam}$). The cantilevered beam was modeled based on linear one-dimensional finite elements with 12 degrees of freedom each. Euler Bernoulli beam theory was used to deal with bending. Using the Euler Lagrange equation with a rotation matrix based on Tait–Bryan angles and using Taylor series for linearization, the mass matrix (\autoref{eq:fea:Mtop}), stiffness matrix (\autoref{eq:fea:K_m_ecc}), and gravitational force (\autoref{eq:fea:Fstat}) of the eccentric concentrated top mass were found. For the case study, the numerical values of the top mass matrix are shown in \autoref{eq:fea:Mtop_num}, the numerical values for the stiffness matrix are shown in \autoref{eq:fea:K_m_ecc_num}, and the numerical values for the foundation spring matrix $\mathbf{K_{found}}$ are listed in \autoref{eq:fea:K_fnd}:

\begin{small}
    \begin{equation}
       \mathbf{M}_{top}  = 
        \begin{bmatrix}
         m_{t}  & 0             & 0             & 0                                             & m_{t}\,z                                  & -m_{t}\,y           \\ 
         0             & m_{t}  & 0             & -m_{t}\,z                                 & 0                                             & m_{t}\,x            \\ 
         0             & 0              & m_{t}  & m_{t}\,y                                     & -m_{t}\,x                                    & 0                    \\ 
         0             & -m_{t}\,z & m_{t}\,y     & I_{c}+\left({z}^2+{y   }^2\right)\,m_{t} & -m_{t}\,x   \,y                             & -m_{t}\,x   \,z \\ 
         m_{t}\,z  & 0             & -m_{t}\,x    & -m_{t}\,x   \,y                             & I_{c}+\left({z}^2+{x   }^2\right)\,m_{t} & -m_{t}\,y   \,z \\
         -m_{t}\,y    & m_{t}\,x     & 0             & -m_{t}\,x   \,z                          & -m_{t}\,y   \,z                          & I_{c}+\left({y   }^2+{x   }^2\right)\,m_{t} 
        \end{bmatrix},
        \label{eq:fea:Mtop}
    \end{equation}
\end{small}

 \begin{small}
    \begin{equation}
       \mathbf{M}_{top}  = 10^{4}
        \begin{bmatrix}
1.7 & 0 & 0 & 0 & 0.2 & 0\\
0 & 1.7 & 0 & -0.2 & 0 & 1.4\\
0 & 0 & 1.7 & 0 & -1.4 & 0\\
0 & -0.2 & 0 & 4.2 & 0 & -0.6\\
0.2 & 0 & -1.4 & 0 & 8.2 & 0\\
0 & 1.4 & 0 & -0.6 & 0 & 8.0
        \end{bmatrix},
        \label{eq:fea:Mtop_num}
    \end{equation}
\end{small}

% \begin{small}
%     \begin{equation}
%         \mathbf{K}_{m_t} =
%         \begin{bmatrix}
%       0 & 0 & 0 &         0        &        0         & 0 \\ 
%       0 & 0 & 0 &         0        &        0         & 0 \\ 
%       0 & 0 & 0 &         0        &        0         & 0 \\ 
%       0 & 0 & 0 & -g\,m_{t}\,z     &        0         & 0 \\ 
%       0 & 0 & 0 &         0        & -g\,m_{t}\,z     & 0 \\ 
%       0 & 0 & 0 &         0        &        0         & 0
%         \end{bmatrix}
%         \label{eq:fea:K_m_ecc}
%     \end{equation}
% \end{small}

\begin{small}
    \begin{equation}
        \mathbf{K}_{m_t} = \textnormal{diag}
        \begin{bmatrix}
			0 & 0 & 0 & -g\,m_{t}\,z  & -g\,m_{t}\,z   & 0
        \end{bmatrix},
        \label{eq:fea:K_m_ecc}
    \end{equation}
\end{small}

\begin{small}
    \begin{equation}
        \mathbf{K}_{m_t} = 10^{4} \textnormal{diag}
        \begin{bmatrix}
			0 & 0 & 0 & -1.9  & -1.9   & 0
        \end{bmatrix},
        \label{eq:fea:K_m_ecc_num}
    \end{equation}
\end{small}

\begin{small}
    \begin{equation}
        \mathbf{f}_{g,m_t} = 
        \begin{bmatrix}
         0 &  0 &  -g(\,m_{t}+m_{t}) &  -g\,m_{t}\,y    &  g\,m_{t}\,x_{e}  &  0
        \end{bmatrix} ^{T},
        \label{eq:fea:Fstat}
    \end{equation}
\end{small}

% \begin{small}
%     \begin{equation}
%         \mathbf{K}_{fnd} =
%         \begin{bmatrix}
%          k_{x} &     0 & 0     & 0         & 0           & 0         \\ 
%          0     & k_{y} & 0     & 0         & 0           & 0         \\
%          0     & 0     & k_{z} & 0         & 0           & 0         \\
%          0     & 0     & 0     & k_{\phi } & 0           & 0         \\
%          0     & 0     & 0     & 0         & k_{\theta } & 0         \\ 
%          0     & 0     & 0     & 0         & 0           & k_{\psi } 
%         \end{bmatrix}
%         = 10^{11}
%         \begin{bmatrix}
%          1.5 & 0 & 0 & 0 & 0 & 0\\ 
%          0 &  1.5 & 0 & 0 & 0 & 0\\ 
%          0 & 0 & 0.44 & 0 & 0 & 0\\ 
%          0 & 0 & 0 & 0.49 & 0 & 0\\ 
%          0 & 0 & 0 & 0 & 0.49 & 0\\ 
%          0 & 0 & 0 & 0 & 0 &  0.094 
%         \end{bmatrix}
%         \label{eq:fea:K_fnd}
%     \end{equation}
% \end{small}

% \begin{small}
%     \begin{equation}
%   \mathbf{K}_{found.} = diag
%   \begin{bmatrix}
% 		k_{x} & k_{y} & k_{z} & k_{\phi} & k_{\theta} & k_{\psi } 
%   \end{bmatrix}
%   = 10^{11}\, diag
%   \begin{bmatrix}
%     1.5 & 1.5 & 0.44 &  0.49 &  0.49 & 0.094 
%   \end{bmatrix}
%   \label{eq:fea:K_fnd}
%     \end{equation}
% \end{small}

\begin{small}
    \begin{equation}
   \mathbf{K}_{found}  = 10^{11}\, \textnormal{diag}
   \begin{bmatrix}
    2.44  &   2.44 &    1.48 &    0.41   &  0.41 &     0.31
   \end{bmatrix},
   \label{eq:fea:K_fnd}
    \end{equation}
\end{small}

where $\textnormal{diag[]}$ denotes the diagonal elements of $6\times6$ matrices where non-diagonal elements are 0.

As loads, gravitational force and an initial deformation of 0.71 m due to a horizontal force applied at the center of the top of the cantilevered beam were considered. To account for numerical diffusivity imposed by linearization of the numerical model, small initial deformations to induce instabilities were applied.

\subsection{Simulation Results}

\begin{figure}
    \centering
    \includegraphics[width=1\textwidth]{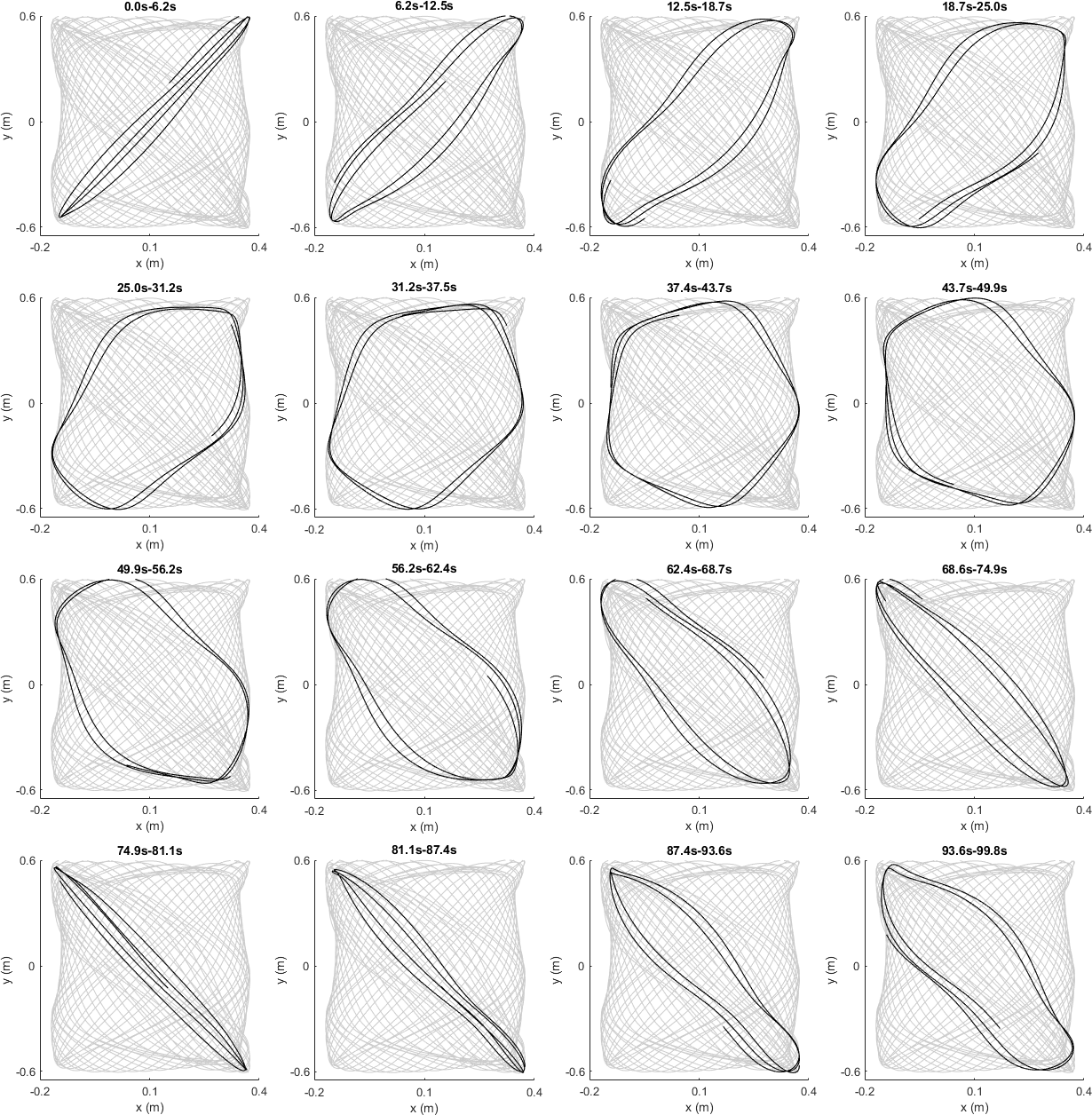}
    \caption{Formation of orbits as observed in the finite element model. The titles of the subfigures denote the time interval of the highlighted orbits.}
    \label{fig:fea:orbit_formation}
\end{figure}

\begin{figure}
    \centering
    \includegraphics[width=1\textwidth]{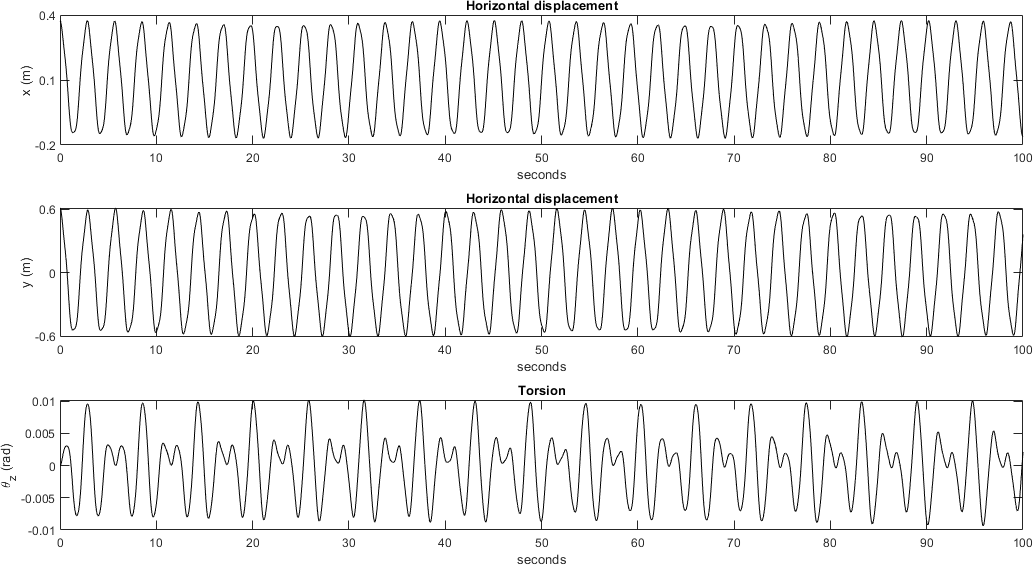}
    \caption{Selected interval of fore-aft (x), side-side (y) and torsion angle $\Theta_z$ time series from the finite element model.}
    \label{fig:fea:simres:timeseries}
\end{figure}

The response to gravitational force and an initial deformation of 0.71 m due to a horizontal force applied at the top of the  cantilevered beam with an angle of 66\textdegree ~with respect to the eccentricity of the top mass is presented in  \autoref{fig:fea:orbit_formation} and \autoref{fig:fea:simres:timeseries}. The pattern found in the simulations (\autoref{fig:fea:orbit_formation}) appear to be quite similar to the pattern found in the experiments, as presented in  \autoref{fig:high-mass}. As in the experiment, torsion around the beam's main axis, albeit small in value, coincided with the presence of orbits. If, however, the initial displacement was in line with the eccentric mass or the eccentric mass was removed completely, no orbits appeared. 

\clearpage

\section{A Vibration Model with three Degrees of Freedom}
\label{sec:3dof}

A common approach to study vibrations is to reduce the real system into a simplified model that consists of discrete components. In this approach, one aims to describe the real system as simple as possible to generate the main dynamics of the real system. Many vibration phenomena can be sufficiently described with systems with only one or two degrees of freedom. Here, we present a model with three degrees of freedom that seems to capture important dynamics of the real system.
\par 

\subsection{Model Derivation}

\begin{figure}
    \centering
    \includegraphics[scale=0.8]{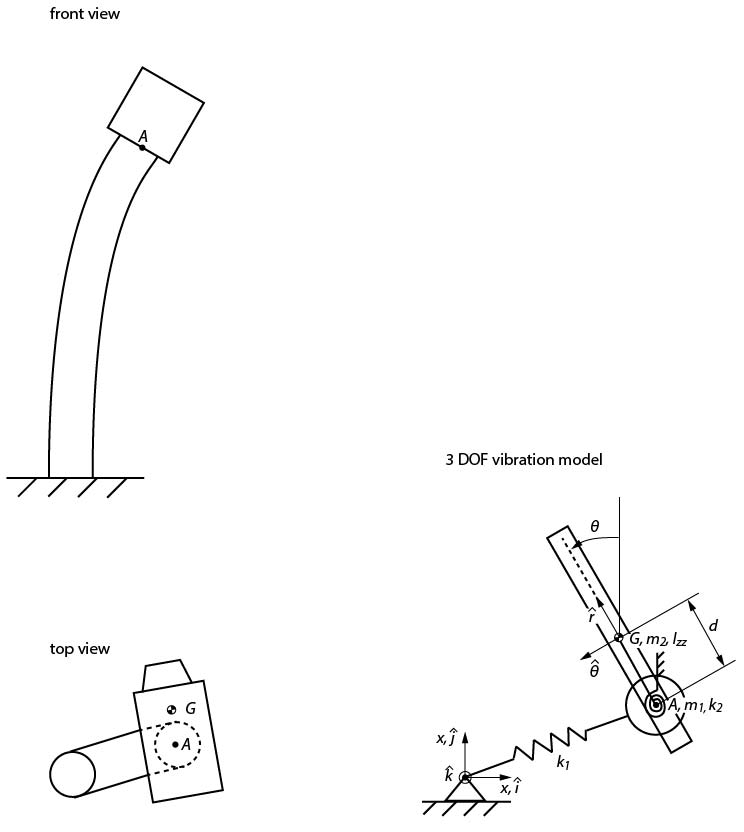}
    \caption{A vibration model with three degrees of freedom (DOF).}
    \label{fig:3dof-system}
\end{figure}

We model the real system in two dimensions (\autoref{fig:3dof-system}). The tower's bending is modeled as a linear spring, which is connected to a fixed point that represents the tower's foundation, and to a first body that represents the tower's top section. The tower's torsion is modeled with a torsional spring that connects a second body with a fixed angle in the inertial reference frame. This second body is pinned to the first body such that it translates with the first body, but rotates freely. Consequently, the system's three degrees of freedom are: the first body's position in $x$ and $y$ direction and the second body's orientation $\theta$.
\par 
By applying the physical laws of conservation of linear and angular momentum, we derived the equations of motions for the system. The system's equations of motion read:
\begin{equation}
    (m_1 + m_2) \ddot{x} - \cos(\theta) m_2 d \ddot{\theta} + \sin(\theta) m_2 d \dot{\theta}^2 + k_1 x = 0,\label{eq:eom-x}
\end{equation}
\begin{equation}
    (m_1 + m_2) \ddot{y} - \sin(\theta) m_2 d \ddot{\theta} - \cos(\theta) m_2 d \dot{\theta}^2 + k_1 y = 0,\label{eq:eom-y}
\end{equation}
\begin{equation}
    (I_{zz} + m_2 d^2)\ddot{\theta} - \cos(\theta) m_2 d \ddot{x} - \sin(\theta)m_2 d \ddot{y} + k_2 \theta = 0,\label{eq:eom-theta}
\end{equation}
where $m_1$ and $m_2$ are the masses of the two bodies, $I_{zz}$ is the second body's moment of inertia, $d$ is the distance between them, $k_1$ is the stiffness of the spring that connects the first body with the origin, $k_2$ is the torsional stiffness of the spring between the first and second body, $x$ and $y$ denote the position of the first body's center and $\theta$ denotes the orientation of the second body (\autoref{fig:3dof-system}). Dots are used to denote time derivatives.

The equations can be arranged to have only acceleration on the left hand side:
\begin{equation}
    \ddot{x} = \frac{1}{m_1 + m_2} \left( \cos(\theta) m_2 d \ddot{\theta} - \sin(\theta) m_2 d \dot{\theta}^2 - k_1 x \right),\label{eq:eom2-x}
\end{equation}
\begin{equation}
   \ddot{y} = \frac{1}{m_1 + m_2} \left(\sin(\theta) m_2 d \ddot{\theta} + \cos(\theta) m_2 d \dot{\theta}^2 - k_1 y \right),\label{eq:eom2-y}
\end{equation}
\begin{equation}
    \ddot{\theta} = \frac{1}{I_{zz} + m_2 d^2} \left(\cos(\theta) m_2 d \ddot{x} + \sin(\theta)m_2 d \ddot{y} - k_2 \theta \right). \label{eq:eom2-theta}
\end{equation}

Several terms couple translation ($x$ and $y$) with rotation ($\theta$): for example, the acceleration in the $x$ direction, $\ddot{x}$, is equal to terms that contain $\dot{\theta}$ and $\ddot{\theta}$ (\autoref{eq:eom2-x}). Such coupling terms are necessary to appear in the equation as otherwise orbits would not change direction (for example, a mass that is initially displaced would indefinitely move in a straight line or a mass that is in an initial orbit would stay in it indefinitely).
\par 
The translational acceleration ($\ddot{x}$ and $\ddot{y}$ in Equations \ref{eq:eom2-x} and \ref{eq:eom2-x}) is affected by the second mass' centrifugal and Euler forces. Centrifugal force acts in the $-\hat{r}$ direction (a coordinate system is shown in \autoref{fig:3dof-system}) and reads, 
\begin{equation}
    m_2 d \dot{\theta}^2,
\end{equation}
and Euler force acts in the $\hat{\theta}$ direction and reads
\begin{equation}
    m_2 d \ddot{\theta}.
\end{equation}
The equation for $\ddot{\theta}$ (\autoref{eq:eom2-theta}) was derived by applying conservation of angular momentum around the center of mass $G$. The force that acts at the location where the second body is pinned to the first body depends on the first body's acceleration in $x$ and $y$ direction such that there are also coupling terms with $\ddot{x}$ and $\ddot{y}$ in the equation for $\ddot{\theta}$.
\par 
As expected, the equations also show that if $d=0$ the coupling terms disappear:
\begin{equation}
    (m_1 + m_2) \ddot{x} = - k_1 x,
\end{equation}
\begin{equation}
   (m_1 + m_2) \ddot{y} = - k_1 y,
\end{equation}
\begin{equation}
    I_{zz}\ddot{\theta} = - k_2 \theta.
\end{equation}

\subsection{Model Results}

\begin{figure}
    \centering
    \includegraphics[width=1\textwidth]{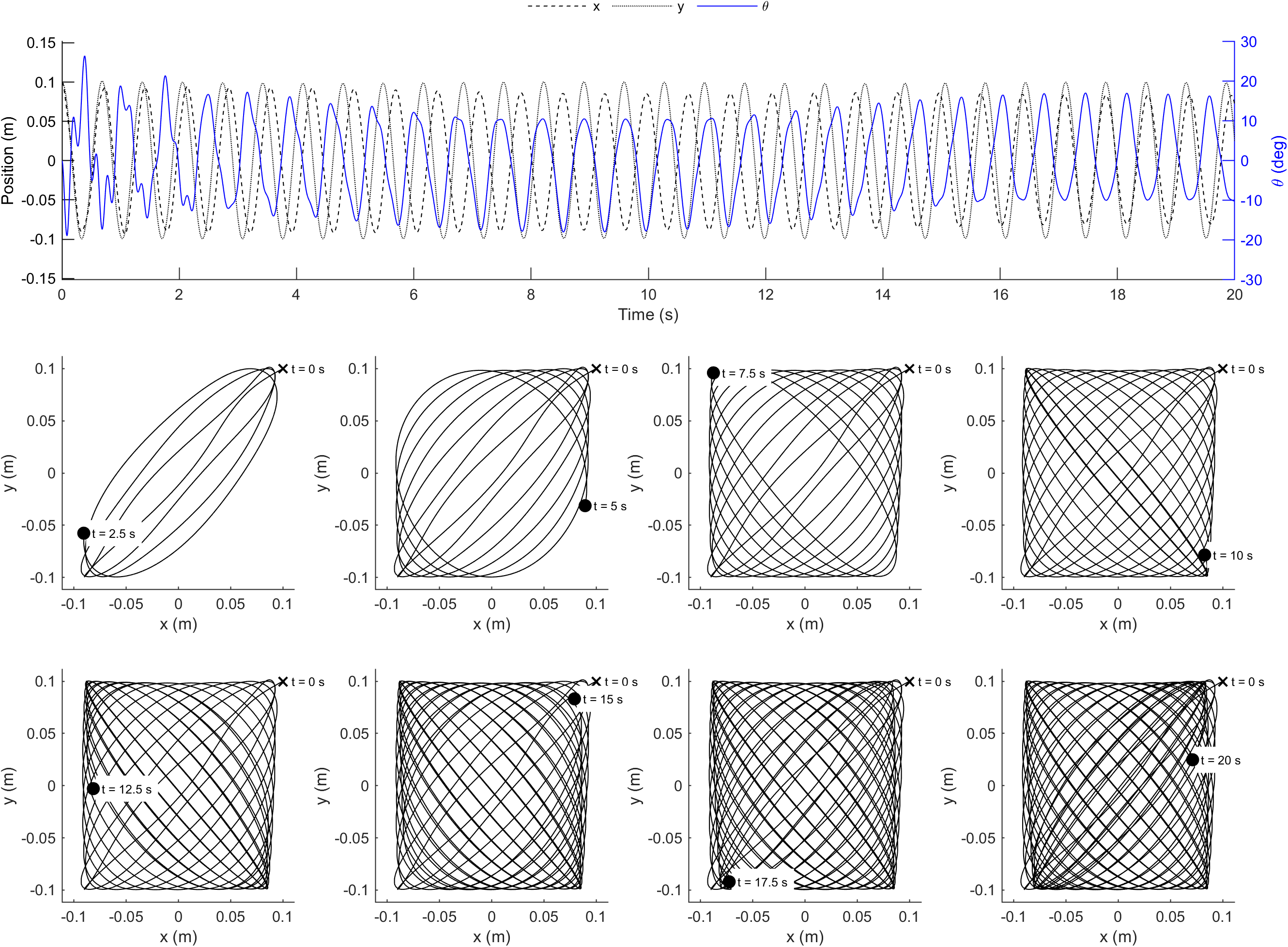}
    \caption{Kinematics of the 3DOF model in the table top configuration (\autoref{tab:3dof-variable-values}).}
    \label{fig:3dof-tabletop}
\end{figure}

\begin{figure}
    \centering
    \includegraphics[width=1\textwidth]{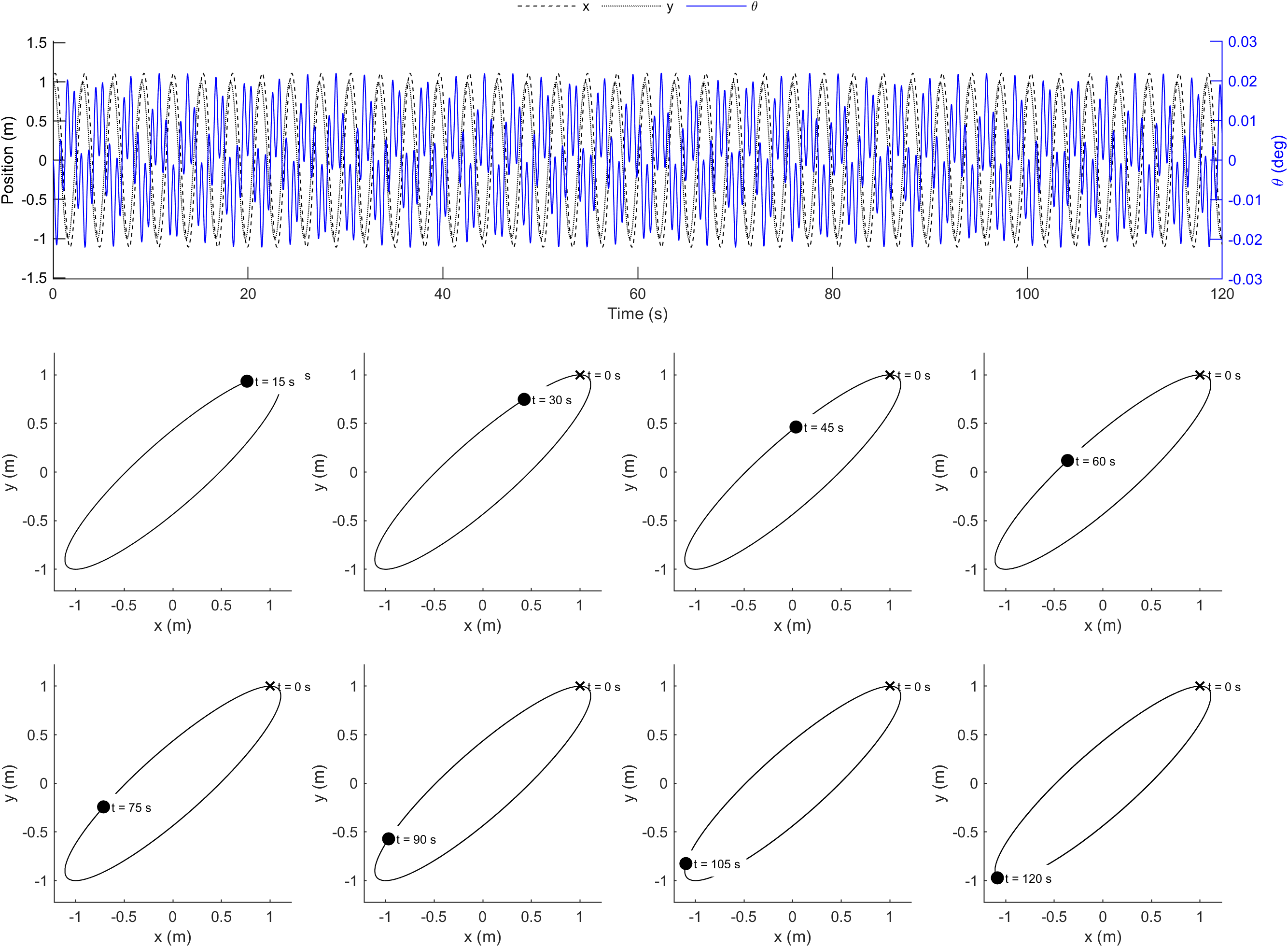}
    \caption{Kinematics of the 3DOF model in a typical wind turbine configuration (\autoref{tab:3dof-variable-values}).}
    \label{fig:3dof-turbine-typical}
\end{figure}

\begin{figure}
    \centering
    \includegraphics[width=1\textwidth]{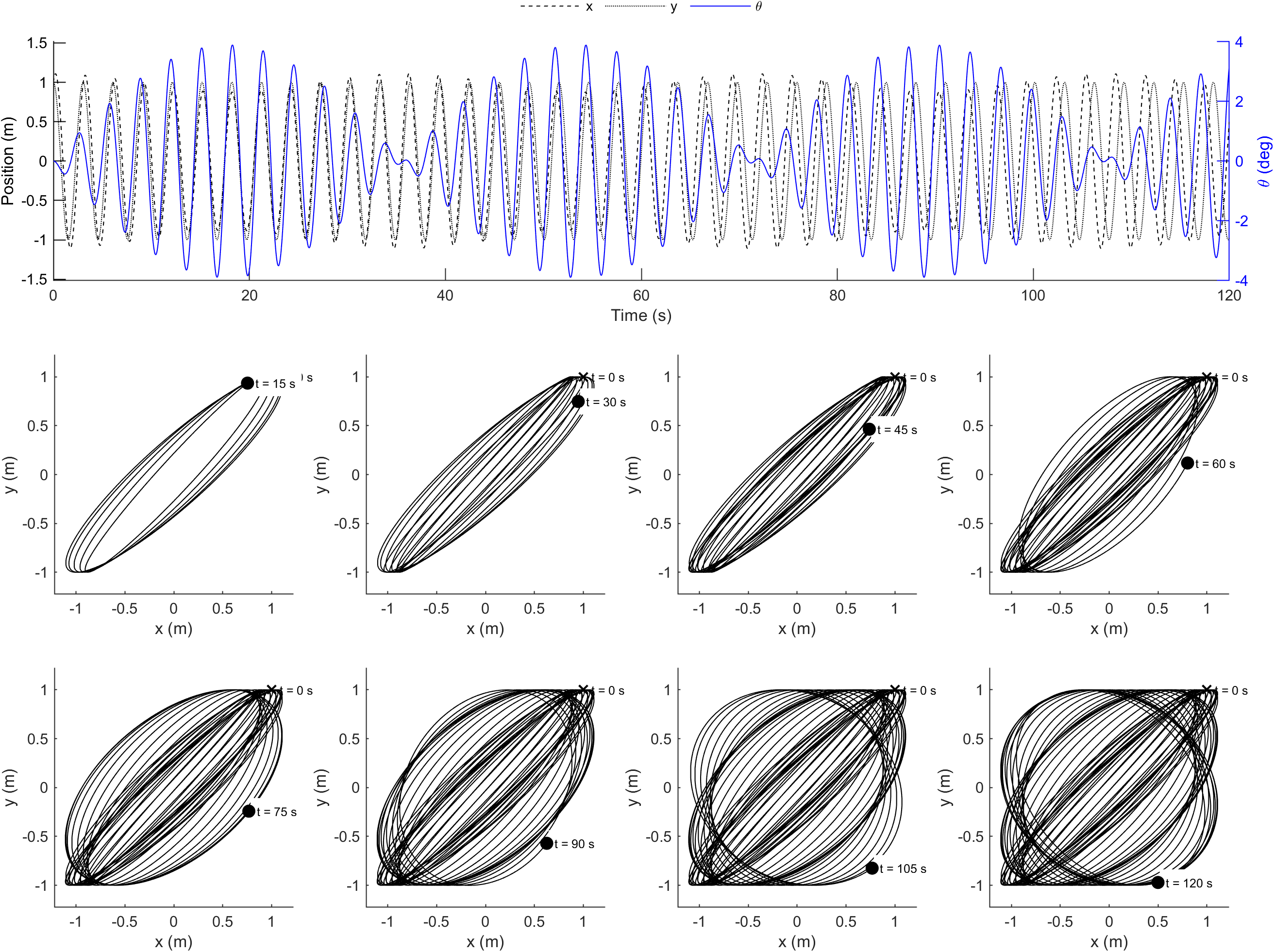}
    \caption{Kinematics of the 3DOF model in a unfavorable wind turbine configuration (\autoref{tab:3dof-variable-values}).}
    \label{fig:3dof-turbine-unfavorable}
\end{figure}

With parameter values similar as in the table top experiment, the vibration model leads to similar motions: orbits with changing major axis direction appear (\autoref{fig:3dof-tabletop}). With parameters values that we consider similar to a typical modern offshore wind turbine in the hammerhead configuration (tower and nacelle installed, but not the blades yet), orbit direction changes do not occur. If the torsional stiffness is very high such that torsional motions are very low, the orbit is stable (\autoref{fig:3dof-turbine-typical}; we consider this scenario typical for a modern wind turbine). In a unfavorable wind turbine configuration where torsional stiffness is lower and the nacelle's center of mass is farther away from the foundation's center, orbits that change appear (\autoref{fig:3dof-turbine-unfavorable}). Similarly as in the finite element model, if the initial displacement is exactly in line with the eccentric mass, no orbits appear.

\begin{table}
    \centering
    \begin{tabular}{l l l ll }
    \toprule
         Quantity & Table top experiment & Offshore turbine &  Offshore turbine & Unit \\
         & & (Typical configuration) & (Unfavorable configuration)\\
         \midrule
         $m_1$ & 0.0093 & 300$\times$10$^3$ & 300$\times$10$^3$ & kg\\ 
         $m_2$ & 0.044 & 400$\times$10$^3$ & 400$\times$10$^3$ & kg\\ 
         $k_1$ & 4.5 & 3$\times$10$^6$ & 3$\times$10$^6$ & N\,m$^{-1}$ \\ 
         $k_2$ & 0.001 & 4$\times$10$^9$ & 4$\times$10$^8$ & Nm\,deg$^{-1}$ \\ 
         $d$ & 0.038 & 0.3 & 1.0 & m\\ 
         $I_{zz}$ & 5$\times$10$^{-5}$ & 4$\times$10$^7$ & 8$\times$10$^7$ & kg\,m$^2$ \\
         \\
         Initial condition & \\
         \midrule
         $x(t=0)$ & 0.1 & 1 & 1 & m\\ 
         $y(t=0)$ & 0.1 & 1 & 1 & m\\
         $\dot{x}(t=0)$ & 0 & 1 & 1 & m\,s$^{-1}$ \\
         $\dot{y}(t=0)$ & 0 & 0 & 0 & m\,s$^{-1}$ \\
         \bottomrule
    \end{tabular}
    \caption{Values used in the 3DOF vibration model that roughly correspond to the table top experiment and the offshore measurements. The resulting kinematics are presented in Figures \ref{fig:3dof-tabletop}, \ref{fig:3dof-turbine-typical}, and \ref{fig:3dof-turbine-unfavorable}}
    \label{tab:3dof-variable-values}
\end{table}

\clearpage

\section{Discussion}
\label{sec:discussion}

\subsection{Table-Top Experiment with a Cantilevered Beam}

Even though the table-top experiment was conducted with very limited resources and meant as a demonstrator experiment only, the formation of orbits with changing directions was clearly observable. More so, the reproducibility of the orbits was high, which suggests that the observed dynamics are stable and can be explained by the properties of the structure instead of external forces such as airflow acting on the structure. In future experiments, the influence of initial and boundary conditions of the experiment should be investigated as it stands to reason, that even without an eccentric mass present at the free end of the beam, orbits will appear in an experiment as simple as ours, given the difficulty of aligning the beam with Earth's gravitational vector. If the beam is tilted with respect to gravity, a certain mass eccentricity will always be induced. Furthermore, in a future study, before designing a new experiment, the governing parameters such as the Scrouton number should be derived from offshore wind turbines by applying similitude first and then incorporated into the design. 

\subsection{Finite Element Simulations of a Cantilevered Beam}

With the linear finite element model, similar orbits as observed in the table-top experiment have been found. Differences may be because the experiments were performed for initial deformations exceeding linear bending behavior of the cantilevered beam, which is not accounted for by the finite element model. However, the similarity between the numerical and experimental results suggest, that the non-linear effects play a minor role in the formation of orbits. In future studies, a more realistic model of a partially installed turbine including non-linear bending could be investigated as well as more realistic, stochastic loading to determine the full mechanics of the partially installed turbine.

\subsection{Three Degrees of Freedom Model}

The three degrees of freedom (3DOF) model presented in this paper leads to orbits comparable to the table-top experiment if the parameters of the differential equations are chosen, such that they are close to the parameters of the table-top experiment. However, for parameters resembling a typical offshore wind turbine, orbits were stable and hardly changed direction. This needs to be investigated in the future. Furthermore, damping could be added as additional coupling terms between translation and torsion are to be expected if damping is included. The configuration of the 3DOF model was created as a direct abstraction of the real system. Alternatively, one could derive a simple vibration model by starting from a mathematical description of a cantilevered beam formulation in three dimensions. Then, one would simplify and neglect terms until a model with few degrees of freedom would remain. Such an approach could be used to decrease model complexity in a step-wise process and to investigate individual model assumptions.

\subsection{Implications for the Kinematics of Partially Installed Wind Turbines}

So far, a coupling mechanism between translational vibrations sufficient to explain the observed orbits in offshore wind turbines undergoing installation is missing. The torsional coupling presented here seems to be a valid candidate; however, given the nature of the governing differential equations, it is to be expected that the torsional coupling itself is a highly complex and possibly chaotic effect. More studies, with parameters closer to real turbines, need to be investigated rigorously. It seems, nevertheless, likely, that some degree of the torsional coupling described here is present in offshore wind turbines as their nacelles are notoriously nose heavy to counter rotor thrust when operational. 

Combining torsional coupling and the stochastic loads inflicted upon the structure by wind and waves could be a sufficient explanation for the formation of orbits and their erratic kinematics during single blade installation.

\section*{Data availability}
 All data of the table-top experiment and the corresponding code is available on github: \url{https://github.com/k323r/2021_preprint_eccentric-mass}. Similarly, the repository contains code that was used to simulate the dynamics of the 3DOF model.
 
\bibliographystyle{rusnat}
\bibliography{references}  %%% Uncomment this line and comment out the ``thebibliography'' section 

\end{document}